 \documentclass[nohyper,12pt,letterpaper]{JHEP3}
 \usepackage{epsfig}
 
 \def \eqn#1#2{\begin{equation}#2\label{#1}\end{equation}}

 \title{Towards a quantum theory of de Sitter space}
 \author{Tom\,Banks\\
 Department of Physics and SCIPP\\
 University of California, Santa Cruz, CA 95064, USA\\
 E-mail: \email{banks@scipp.ucsc.edu}\\
 {\it and}\\
 Department of Physics and NHETC, Rutgers University, USA\\
 Piscataway, NJ 08540}
 \author{Bartomeu\,Fiol\\
Institute for Theoretical Physics, University of Amsterdam\\
1018 XE Amsterdam, The Netherlands\\
E-mail: \email{bfiol@science.uva.nl}}
 \author{Alexander\, Morisse\\
 Department of Physics and SCIPP\\
 University of California, Santa Cruz, CA 95064, USA\\
 E-mail: \email{amorisse@physics.ucsc.edu}}

 \abstract{We describe progress towards constructing a quantum theory of de Sitter space in
 four dimensions.  In particular we indicate how both particle states and Schwarzschild de Sitter
 black holes can arise as excitations in a theory of a finite number of fermionic oscillators.
 The results about particle states depend on a conjecture about algebras of Grassmann variables,
 which we state, but do not prove.}

 \received{} \accepted{}
 \preprint{hep-th{0609062}\\\\RUNHETC-06-21, SCIPP-06-11\\ ITFA-2006-31 \\}
\begin{document}
\section{\bf Introduction}

The observed acceleration of cosmic expansion is most simply
explained by positing a small positive cosmological constant (c.c.)
in the low energy effective Lagrangian for the metric of four
dimensional space.  The existence of a positive cosmological
constant raises a host of problems for theoretical high energy
physics. One would like to understand the magnitude of this
parameter, and the nature of quantum observables in the
asymptotically de Sitter (dS) space-time that results.

With regard to the first question, Weinberg's {\it galactothropic}
bound on the cosmological constant\cite{wein} seems to provide a
satisfactory answer to the question both of the absolute magnitude
and the ratio of the c.c. to the current matter density.   This
assumes that one has a theory in which the dark matter density
$\rho_0$ and the strength of primordial fluctuations $Q$ are fixed
as one varies the c.c.   or that $Q$ and $\rho_0$ vary with
$\Lambda$ in such a way as to take on their real world values at the
real world value of $\Lambda$.   Theories in which all three
parameters are independent random variables are somewhat less
attractive, but can still account for a reduction of the \lq\lq
expected" value of the c.c. by many orders of
magnitude\cite{wiseetal}.

There are two classes of meta-physical theories which are based on
plausible dynamics, and could give rise to a plethora of universes
with different values of the c.c.    We call them meta-physical
because, at least with current understanding, there is no way to
make observations on the alternative universes.   The more famous of
the two is called the String Landscape\cite{landscape}, though much
of it is based on ideas that are more than 20 years old and had no
connection with string theory. Holographic cosmology\cite{holocosm}
presents a picture of a metaverse consisting of many asymptotically
dS space-times embedded in a dense black hole fluid.   In the
Landscape, all parameters of low energy physics vary as we jump from
universe to universe, while in holographic cosmology it is plausible
that only the c.c. and things which depend on it in the limit when
it is small, vary.  There is as yet no mathematical formulation of
what the Landscape is, though there have been some interesting
attempts to construct one\cite{lennyetal}.   The holographic
cosmology of a defect free dense black hole fluid has been
mathematically formulated\cite{holocosm}, but not the theory of the
asymptotically dS defects in the fluid.

Metaphysical theories are useful, if at all, only for understanding
values of low energy parameters\footnote{Here we conform to the
conventional and incorrect language which identifies the c.c. as a
low energy parameter.   In all examples we understand, it is in fact
related to the physics of the highest energy states in the theory,
which are large black holes.}, which cannot be explained by ordinary
dynamical mechanisms.   To a large extent, once the parameters are
chosen the metaphysical theory is of little practical relevance.
This is particularly true in the holographic cosmology approach,
where the structure of the theory dictates that dynamics inside each
asymptotically dS bubble is independent of the outside.

{}From this point of view at least, the task of theoretical physics
is to construct a quantum theory of a single asymptotically dS
universe.  It may be that such a theory will also be useful as an
approximation to a future mathematical theory of the Landscape.  At
any rate, this is the task we will take up in the present paper. In
fact, we will take up a somewhat more modest task.  In the mid 80s,
the focus of string theorists was on finding a Poincar\'e invariant
description of the real world.   This was not because string
theorists were ignorant of the existence of cosmology, but because
they imagined that the laws of particle scattering were, to a good
approximation independent of both the asymptotic past and future.
This project failed because no one has ever found a consistent model
of quantum gravity which is Poincar\'e invariant in d flat spacetime
dimensions without being super-Poincar\'e invariant.   Nor have we
found asymptotically AdS theories with radius large compared to the
string scale in which SUSY is broken in most of the volume of AdS.

Our own take on this failure is that it indicates a close connection
between the SUSY breaking we observe in the real world, and the fact
that we have a positive c.c.\cite{tbfolly} .   The analog of the old
string theory program is to construct a theory in eternal dS space,
which will contain an approximation to particle physics where SUSY
breaking is evident, but the time dependence of cosmology is
neglected.  If you are with us so far, we can proceed to the
construction of the theory.

\section{General rules of holographic space-time}

In the course of constructing a holographic cosmology, the authors
of \cite{holocosm} also invented a general formalism for the
holographic description of space-time. It is motivated by simple
kinematic considerations about the nature of observers in quantum
mechanics, the causal structure of Lorentzian space-times, and the
covariant entropy bound\cite{fsb}.

In quantum mechanics, {\it an observer} is a large quantum system
with many semiclassical observables.   The only way we know how to
construct mathematical models of such systems is to use the rules of
(possibly cut off) quantum field theory.    Indeed, a field theory
in finite volume is such an observer.   Averages of local fields
over significant fractions of the volume have very small quantum
fluctuations.   The tunneling amplitudes between states with
different values of these macroscopic pointer variables go like $e^{
- c V M^d}$ where $M$ is the cutoff momentum scale and $d$ the
dimension of the world volume of the field theory. The precise
observations of the mathematical formalism of quantum mechanics, in
which we imagine that observables of a small system can be measured
with arbitrary precision, are well approximated by machines which
follow the rules of quantum field theory, for $V M^d$ a few orders
of magnitude.   In realistic laboratory situations this number is of
order $V M^d \sim 10^{23}$.

It is an experimental fact, and follows from the rules of quantum
field theory, that any such observer has a large mass, and will
follow a time-like trajectory in $D$ dimensional space-time.   The
rules of holographic space-time are constructed in order to describe
the observations made by these time-like observers in a way that is
compatible with the holographic principle.   Associated with any
segment of a future directed time-like trajectory, going from a
point P to a point Q in its future, there is a causal diamond,
consisting of the intersection of the interior of the backward
light-cone of Q with that of the forward like cone of P.   The
boundary of this diamond is a null surface, and the maximal area
space-like $D - 2$ submanifold on this surface is called the
holographic screen of the causal diamond.   The covariant entropy
bound bounds the entropy flowing out through the future boundary of
the diamond, by one quarter of the area of this screen, in Planck
units.   We can reconstruct the time-like trajectory from a nested
sequence of causal diamonds.

In quantum mechanics entropy always refers to a particular density
matrix, and Fischler and Banks argued that the only natural density
matrix to choose in implementing this principle for a generic
space-time is the maximally uncertain density matrix, proportional
to $1$.   Alternatives, like a thermal density matrix, require a
preferred definition of the Hamiltonian, which is anathema in a
generally covariant theory.   With this choice, the entropy referred
to in the covariant bound is the logarithm of the dimension of the
Hilbert space associated with the diamond.

There is a natural way to construct a finite dimensional Hilbert
space from classical constructs associated with the diamond.
Consider a small area, or pixel, on the holographic screen.   The
screen lies on a null surface and, on the center of this pixel there
is an orthogonal null ray which penetrates the pixel.   Actually
there is an ambiguity here corresponding to whether the null ray is
ingoing or outgoing.   We should use both, but imagine that dynamics
relates the incoming to outgoing rays, so we will describe only
variables associated with the outgoing ones.   The null direction,
and the orientation of the pixel, which is a bounded region of a
space-like $D - 2$ plane orthogonal to the null direction, are both
encoded in a pure spinor, a solution of the Cartan-Penrose equation:

\eqn{cpeqn}{\bar{\psi} \gamma^{\mu} \psi \gamma_{\mu} \psi = 0.}

The independent real solutions to this equation are quantized
according to \eqn{acr}{ [S_a , S_b ]_+ = \delta_{ab}.} $S_a$, or
possibly complex linear combinations of them (depending on $D$)
transform as spinors in the tangent space of the holographic screen.
More generally, we can and should enlarge the algebra of pixel
operators to include information about compactified spatial
dimensions\cite{vn}.

The quantization of pixel operators defines an area for the pixel,
via the Bekenstein-Hawking relation.   Extending this to all pixels
of all holographic screens of all causal diamonds of all observers,
one would (over) determine the Lorentzian geometry of space-time. A
finite area holographic screen would have an operator algebra
\eqn{scrnopalg}{ [S_a (m) , S_b (n) ]_+ = \delta_{ab} \delta_{mn}, }
where we have exploited a $Z_2$ gauge invariance of the CP equation
and the commutation relations to Klein transform the commutation
relations between independent pixels into anti-commutation
relations.   This $Z_2$ is associated with space-time $( - 1)^ F$
and imposes the spin-statistics connection on our formalism.

The operators $S_a (n)$ should be thought of as a section of the
spinor bundle of the holographic screen.  The fact that they are
finite in number (since the screen has finite area) means that we
have pixelated the screen by replacing its function algebra by a
finite dimensional algebra.

It is extremely interesting that the representation space of the
$S_a (n)$ for fixed $n$ is precisely that of the spin degrees of
freedom of a massless superparticle with fixed momenta.  So the
degrees of freedom of a pixel on the holoscreen are precisely those
of a massless superparticle which penetrates that pixel.

Let us now specialize to the case of four dimensional de Sitter
space.   It is clear that our set-up is observer dependent.  In a
symmetric space we are free to choose special coordinates which
exhibit the symmetry.   In order to describe the holographic screen
of the cosmological horizon of an individual observer we want to
implement the obvious $SU(2)$ symmetry of the screen.   We also have
a special operator corresponding to the generator of motion $H$
along the time-like Killing vector seen by this observer. $H$ will
be the focus of our attention in the next section.

An $SU(2)$ invariant way of pixelating the geometry of the
two-sphere is to introduce the fuzzy sphere algebra (the algebra of
$N \times N$ matrices) as the definition of the topology of the
sphere\footnote{Gelfand's theorem shows us the functorial identity
between the topology of a compact Hausdorff space and its
commutative C* algebra of complex valued functions.  A key idea of
non-commutative geometry is to simply identify a space with every C*
algebra.}.   The spinor bundle over the fuzzy sphere is the module
of rectangular $N\times N + 1 $ matrices, which contains all half
integral spin representations of $SU(2)$ up to $N - {1\over 2}$.

The operator algebra of the cosmological horizon is thus
    \eqn{coshoropalg}{[\psi_i^A , (\psi^{\dagger})^j_B ]_+ =
    \delta_i^j \delta^A_B .}
The dimension of this fermionic Hilbert space is $2^{{N(N + 1)}} $,
which means that in the large $N$ limit we should identify
\eqn{bharea}{4 N^2 {\rm ln}\ 2 = \pi (RM_P)^2, } where $R$ is the dS
radius.  For dS space, we believe the density matrix to be $e^{-
2\pi R H} $, but we will see below that the spectrum of $H$ is such
that this estimate of the relation between $R$ and $N$ is unchanged
for large $N$.

\section{The static hamiltonian and the hamiltonian of Poincar\'e}

The causal diamond of a time-like geodesic observer in dS space is
covered by the static coordinate patch

\eqn{staticpatch}{ds^2 =  - f(r) d\tau^2 +{{dr^2}\over f(r)} + r^2
d\Omega^2 ,} where $$f(r) = 1 - {{r^2}\over {R^2}}.$$ In \cite{tbds}
one of the authors presented a qualitative ansatz for the
Hamiltonian $H$, which generates time evolution with respect to the
static time $\tau$. The claim is that the spectrum of $H$ is bounded
by $ c T_{dS} = {c\over 2\pi R}$, where $c$ is a constant of order
one. The Hilbert space on which $H$ acts has dimension of order
$e^{\pi (RM_P)^2 }$, where Newton's constant, $G = M_P^{-2} $.   The
spectral density of $H$ is thus $e^{ - \pi (RM_P)^2}$.

Consider, for fixed c, a random Hamiltonian with these properties,
chosen from any smooth probability distribution, and a fixed initial
state, $| \psi >$ in the Hilbert space.   Since $H$ is random, time
averaged correlation functions in the state $| \psi >$ will approach
thermal equilibrium, with an equilibrium temperature $T = k_{\psi}
T_{dS} $.   We conjecture that, for large $R$, and generic $| \psi >
$ (chosen from a uniform probability distribution on the unit sphere
in Hilbert space), $k_{\psi}$ will approach some average value
$\bar{k}_{\psi}$ . By adjusting $c$ we can choose $\bar{k}_{\psi} =
1$.    Thus we claim that the spectral characteristics assumed for
$H$ can {\it explain} why dS space is a thermal system with a unique
temperature (see appendix for details).

This picture of the spectrum of $H$ is supported by two other
observations about dS space.   The most striking is the formula for
Coleman-De Luccia (CDL) tunneling probabilities between two
different dS vacua.   These satisfy a law of detailed balance
consistent with the picture of dS space as a quantum system with a
finite number of states, {\it but with the free energy replaced by
the entropy}.   This makes sense for a thermal density matrix if the
spectrum of the Hamiltonian is bounded by something close to the
temperature.

The second piece of evidence is that every localized object in de
Sitter space decays to the vacuum.   Classically the vacuum has zero
energy and it makes sense to say that the real quantum vacuum is an
ensemble of states with energies below $T_{dS}$ (which vanishes in
the limit $R \rightarrow \infty$).   A black hole has a large
classical mass, and it is inconsistent with energy conservation to
say that it decays to a state with small energy.   We conclude that
the black hole mass cannot be close to an eigenvalue of $H$.

The aim of \cite{tbds} was to come up with a quantum model that
accounts for all of the things we think we know about dS space from
the approximation called quantum field theory in curved space-time.
That approximation indeed finds a thermal density matrix at the dS
temperature, but with a Hamiltonian, which we will call $P_0$, whose
eigenvalues include particle masses. QFTCST theorists think of $P_0$
as the generator of static time translations, but they also claim
that $P_0$ approaches the ordinary Hamiltonian of a Poincar\'e
invariant theory as $R \rightarrow \infty$.   This is inconsistent
with our model for the spectrum of $H$.

$P_0$ and $H$ are thus different operators.   $P_0$ is in fact only
an {\it emergent quantity}.   It describes localized excitations in
a given horizon volume, which are unstable to decay (via the true
$H$ dynamics of the system) into the dS vacuum.   It is useful
because the time scales involved in $P_0$ dynamics are short
compared to the decay times of the excitations.   We view the
eigenvalues of $P_0$ to be the proper description of particle masses
and the masses of black holes.

To understand how the thermal ensemble with Hamiltonian $H$ can look
like the thermal ensemble with Hamiltonian $P_0$, we postulate a
peculiar relation between the $P_0$ eigenvalue and the entropy
deficit of the corresponding eigenspace.   The calculations done in
QFTCST are an approximation in which decays of most localized
systems do not occur.   In this approximation we do not really
resolve the spectrum of $H$ and it makes sense to take $H \approx
0$.   In that case the probability in the thermal density matrix of
$H$ for having $P_0$ eigenvalue $E$ is just \eqn{p0prob}{{\rm Tr}\
e_{E} e^{ - \pi (RM_P)^2} , } where $e_E$ is the projector on the
eigenspace with $P_0 = E$. This will reproduce the  probability
computed from the thermal density matrix for $P_0$ if \eqn{trp0}{
{\rm Tr}\ e_{E} = {{e^{ \pi (RM_P)^2 - E/T_{dS}}}\over {{\rm Tr}\
e^{- P_0 / T_{dS}}}} .}  This relation between the Poincar\'e
eigenvalue and the entropy deficit is valid to leading order in ${M
\over {R(M_P)^2}}$ for Schwarzschild de Sitter black holes of mass
$M$.    We view this as another piece of semi-classical evidence for
the picture advanced here.

The commutator between $H$ and $P_0$ has the form \eqn{hp0}{[H , P_0
] = \sum (E_j - E_i) H_{ij} ,} where $H_{ij}$ is the rectangular
block of $H$ with rows in the $i$th and columns in the $j$th
eigenspace of $P_0$.   The individual matrix elements in any of the
$H_{ij}$ are bounded by ${c\over 2\pi R}$, but most of them are much
smaller than this.   In particular the QFTCST claim that the effects
of $H$ dynamics look like thermal fluctuations in the thermal $P_0$
ensemble, tell us that matrix elements connecting huge $E_i - E_j$
difference are exponentially suppressed.   The spectrum of $P_0$ is
bounded by the Nariai black hole mass, which is of order $R
M_P^2$\footnote{Note that black holes are not eigenstates of $P_0$.
However, this is a bound on the energy of the decay products.   For
black hole masses much smaller than the Nariai mass these decay
products can be captured by the static observer and might have a
lifetime much longer than that of the black hole.  The decay of this
bound but not gravitationally collapsed system back to the dS vacuum
is probably not encoded in the operator $P_0$, but only in $H$. The
captured decay products could be in eigenstates of $P_0$}.

In \cite{tbds} one of the authors postulated the commutation
relation between these two operators to be a finite dimensional
approximation to $[H, P_0 ] \sim {1\over R} P_0 $.   This was
motivated by the way the asymptotic Killing vectors of Minkowski
space act on the dS horizon. The general considerations above show
that the commutator is small but do not point to this specific form.

\section{Black holes from fermionic matrices}

This section is meant to replace a somewhat confused discussion in
\cite{tbds}. The metric of the Schwarzschild dS black hole has the
same form as the static patch metric with $f(r) \rightarrow (1 -
{{2M}\over {r M_P^2}} - {{r^2}\over {R^2}})$. This has two horizons,
which are at the positive roots of
$$r^3 - r R^2 + {{2M R^2}\over {M_P^2}} = 0 .$$  We write this as
\eqn{R}{R^2 = R_+^2 + R_-^2 + R_+ R_- ,} \eqn{M}{2M R^2 = R_+ R_-
(R_+ + R_-)M_P^2 .}   The entropy deficit of the black hole state,
taking into account both cosmological and black hole horizons, is
\eqn{entdef}{\Delta S = \pi R_+ R_- M_P^2 .}

We match the entropy of our fermionic Hilbert space to that of dS
space by \eqn{entmatch}{\pi (RM_P )^2 = 4 N^2 {\rm ln}\ 2 .}  All
such formulae are to be understood only in the large $N$ limit.  In
order to find candidate black hole states, we choose two integers
related to $R_{\pm}$ by the same formula

\eqn{entmatch}{\pi (R_{\pm} M_P )^2 = 4 N_{\pm}^2 {\rm ln}\ 2 .}

We do this by choosing an integer $N_- \leq {1\over\sqrt{3}} N,$ and
defining $N_+$ to be the closest integer to the solution of
\eqn{neqn}{N^2 = N_+^2 + N_-^2 + N_+ N_- ,} satisfying the
constraint \eqn{nconst}{N_+ \geq N_- .}   Now choose $N_-$ rows and
$N_+$ columns of the fermionic matrix $\psi_i^A$ and define the
black hole states with Schwarzschild radius $N_-$ to be those
annihilated by $\psi_i^A$ for the chosen rows and columns.   The
reader is encouraged to think of the choice of a particular set of
$N_-$ rows and $N_+$ columns as analogous to the choice of a
particular static coordinate system.   Note for example that as
$N_-^2$ gets large, and approaches its maximum value, ${N^2 \over
3}$, one cannot independently choose to construct black hole states
for arbitrary choices of rows and columns.   We will have more to
say about the way that different horizon volumes are embedded in the
index space of the fermionic matrices when we discuss particle
states below.

We can reproduce the black hole mass formula \ref{M} by writing
\eqn{p0approx}{P_0 = \sqrt{{{\rm ln} 2}\over {2\pi}} M_P (N^2 -
2{\cal N})\sqrt{N^2 - {\cal N}}.} Here, ${\cal N}$ is the total
fermion number operator. This formula is \lq\lq coordinate
invariant", in the sense that it makes no reference to the
particular choice of rows and columns. The black hole states we have
defined are all eigenstates of this operator, but not all with the
same eigenvalue. However, for large $N$, the average value of $P_0$
in the ensemble of all black hole states is indeed the classical
black hole mass, and the fluctuations in this ensemble go to zero
like a power of $N$. We make the further rule that, when speaking of
a particular horizon volume we only look at states with a particular
choice of $N_-$ rows and $N_+$ columns.

A puzzling feature of the fermionic formulation is that one could
consider similar states with arbitrary choice of $N_+$ independent
of $N_-$.   This is perhaps related to black holes with angular
momentum, but we have not yet studied the angular momentum
properties of these states.

While the operator $P_0$ realizes the relation between entropy and
energy that we described in the previous section, it is far from the
exact Hamiltonian characterizing quantum dS space.  We view it as
the {\it asymptotic darkness}\cite{tbsb} approximation to that
Hamiltonian, in which only black hole spectra are treated, and black
holes are exactly stable.   We will have to modify the Hamiltonian
in order to describe black hole decay, and the particle states they
decay into.   It is to the second part of this task that we now
turn.

\section{Particles from fermionic matrices}

Before embarking on this task, we should recall the extent to which
physics in dS space can be described in terms of particles.  We
begin our analysis, {\it faute de mieux} with quantum field theory,
though we will see below that there is a more elegant description
available.   Our question is: {\it How much of the entropy of dS
space can be understood in terms of particles?} and it was answered
in \cite{nightmare}.   The entropy of quantum field theory is
dominated by high energy states, and the high energy behavior of a
quantum field theory is conformally invariant.   The entropy of a
cutoff conformal field theory in a volume of linear size $R$ scales
like
$$\Lambda_c ^3 R^3, $$ where $\Lambda_c$ is the momentum cutoff.
A typical state in this ensemble has energy of order
$$ \Lambda_c^4 R^3 < M_P^2 R,$$ where the inequality is the
requirement that the Schwarzschild radius of the state be less than
the dS horizon radius.   This implies that the cutoff is very low
$$\Lambda_c < \sqrt{M_P \over R}. $$   One should understand that
this is not the limit on the momentum of individual particles in
isolation, but only of particle belonging to the maximal entropy
ensemble.   Our actual description of particles in dS space will
have the tradeoff between the momenta of individual particles, and
the total allowed number of particles, built in to it. With this
cutoff, the total field theory entropy is of order $(RM_P)^{3/2}$.
In \cite{nightmare} the authors suggested that this counting allows
us to understand the QFTCST picture of dS space as a system which
(at asymptotically late or early times) has an infinite number of
independent horizon volumes, each described by cutoff field theory.
Our counting suggests that the number of independent field theoretic
subsystems is finite, of order $(RM_P)^{1/2} $, but becomes infinite
in the small c.c. limit where QFTCST is supposed to be a good
approximation.

We now proceed to present our model in more detail. The holographic
formulation of quantum gravity in de Sitter space we want to put
forward is expected, in the $\Lambda\rightarrow 0$ limit, to recover
$4d$  ${\cal N}=1$ Super-Poincar\'e physics. At the kinematical
level, we expect to recover something akin to Ashtekar's formalism
for asymptotically flat spacetime at null infinity \cite{ashtekar}.
For $4d$ Minkowski space, null infinity is $(u,\Omega)$, where $u$
is a null coordinate and $\Omega$ parameterizes a two sphere. The
conformal group of the sphere is $SO(1,3)$, which is identified as
the Lorentz transformations.   However, as we will see below, the
kinematical theory of dS space does not lead to a field theoretic
formalism on null infinity.   Rather, in close analogy with Matrix
Theory, we obtain a direct description of multi-particle states from
a theory of matrices.

To set the stage, we discuss the formulation of SUSY algebra and
SUSY multiplets of ${\cal N}=1$ at null infinity. First, we will
show that the 4 independent solutions to the Conformal Killing
spinor equation for $S^2$ provide a realization of the minimal
supersymmetry algebra in 4d for massless multiplets.  The conformal
Killing spinor equation is \eqn{CKS}{D_{\mu} q^{(\alpha )} =
\gamma_i e^i_{\mu} \lambda^{(\alpha)} .}

In the usual angular coordinates on the two sphere, the zweibein has
non-vanishing components \eqn{zwei}{e_{\theta}^1 = 1 , \ \ \
e_{\phi}^2 = {\rm sin} \theta.} In the representation where the two
dimensional Euclidean Dirac matrices are $\sigma_{1,2}$, the spinor
covariant derivatives are \eqn{covder}{D_{\theta} =
\partial_{\theta}, \ \ \ D_{\phi} = \partial_{\phi} - {i\over 2}
\sigma_3 {\rm cos}\theta .}   Four linearly independent solutions of
the CKS equation are

\eqn{cks1}{q^1 = i  \sqrt{1-{\rm cos} \theta} e^{i\phi /2}
\pmatrix{0\cr 1} ,} \eqn{cks2}{q^2 = -i \sqrt{1+{\rm cos} \theta}
e^{-i\phi /2} \pmatrix{0\cr 1} ,} \eqn{cks3}{q^3 = \sqrt{1+{\rm cos}
\theta} e^{i\phi /2} \pmatrix{1\cr 0},} \eqn{cks4}{q^4 =
\sqrt{1-{\rm cos}\theta} e^{-i\phi /2} \pmatrix{1\cr 0} .} These
satisfy \eqn{cksprod}{(q^{\dagger})^{\alpha} q^{\beta} = (\gamma^0
\gamma^{\mu} )^{\alpha\beta} \hat{P}^{\mu} , } with the Weyl
representation of the $SO(1,3)$ Dirac matrices.  Here
\eqn{unmom}{\hat{P}^{\mu} = (1, {\rm sin}\ \theta {\rm cos}\ \phi ,
{\rm sin}\ \theta {\rm sin}\ \phi ,{\rm cos}\ \theta ). } Below we
will argue that $K\times K + 1$  fermionic matrices $\psi_i^A $,
converge as $K$ goes to infinity to operator valued linear
functionals on the space of measurable sections of the spinor bundle
of the two sphere.   For two sections $f_a$, $g_a$, the commutation
relations are
$$[ S[f], S[g] ]_+ = p f_a g_a ,$$ where $p$ is a positive real
number we will explain below.    It follows that
$$[S[q^{\alpha}] , S[q^{\beta}]]_+ = (\gamma^0
\gamma^{\mu})^{\alpha\ beta} P_{\mu} .$$

We now recall that any CPT-invariant massless multiplet of ${\cal
N}=1$ $d=4$ SUSY contains states with 4 helicities:
$m,m-1/2,-m+1/2,-m$. At null infinity, these states are sections
over $S^2$ line bundles with the corresponding charges.  The charge
of a line bundle is one half the power of the positive chirality
spinor bundle which realizes it. This is the description of
asymptotically flat ${\cal N}=1$ $d=4$ kinematic framework we aim to
recover in the $\Lambda \rightarrow 0$ limit of deSitter space.
Coming back to deSitter, at finite N, the $S^2$ sphere at null
infinity is substituted by a fuzzy sphere, and the $S^2$ line bundle
of charge $m$ is substituted by a module, a vector space of $N\times
(N+2m)$ matrices \cite{grosse}. Our main claim is this section is
that we can take blocks of Grassmann variables, such that, in the
limit their size goes to infinity, the 4 sections of a
supermultiplet are recovered\footnote{Our discussion here bears some
relation to that of \cite{parverl}.   We suspect that the relation
between their framework and ours is similar to that between Hilbert
spaces of gauge theories before and after fixing gauge constraints:
in their Hilbert space, with dimension growing like $e^{R^3}$, not
all degrees of freedom are simultaneously physical. After one
chooses
 a holographic screen ("fixes the gauge"), the reduced Hilbert space
 has dimension $e^{R^2}$, as ours has.}.

Our proposal for the chiral multiplet, $m=1/2$, uses the coherent
state representation of the operator algebra of the variables
$\psi_i^A$.  The Hilbert space consists of functions of a Grassmann
variable $z_i^A$, $i=1,\dots, p$, $A=1,\dots,p+1$ as a $p\times
(p+1)$ matrix, with the usual Berezin inner product. As in Matrix
Theory, we mean arbitrary functions of the matrix elements.

Now introduce \eqn{na}{ n^a \equiv z^T J_{p}^a z} and \eqn{Na}{N^a
\equiv z J_{p+1}^a z^T ,} where $J_{2l+1}^a $ are the $2l + 1$
dimensional representation of the angular momentum matrices. The
space of all holomorphic Grassmann functions of $z$ decomposes
into four subspaces, which are of the form \eqn{f1} {z f_1 (n^a )}
$$ f_2 (n^a ) ,$$  $$ f_3 (N^a ) ,$$ $$z^T f_4 (N^a ). $$
In these formula, the functions $f_i$ are {\it matrix polynomials}.
For example, the general form of $f_2$ is $\sum c_{a_1 \ldots a_k}
n^{a_1} \ldots n^{a_k} ,$ where the products are $ p + 1 \times p+1$
matrix products.   Our claim is that every function of the matrix
elements takes one of these four forms.  That is, the matrix
elements of these four kinds of matrix fill out the space of all
functions of the matrix elements of $z_i^A$ in a one to one fashion.

 Note that the differences
between the numbers of rows and columns in these four subspaces are
1,0,0,-1, as appropriate for the modules of the four sections of the
$m=1/2$ supermultiplet. The $f_i $ are all power series in their
respective matrix variables. The power series all truncate, because
$z$ is a Grassmann variable.

Our conjecture is that, one can take the
$p \rightarrow \infty $ limit, in such a way that these four
subspaces of the Hilbert space converge to the space of $L^2$
sections of four line bundles over the two sphere.  This conjecture
is motivated by the way in which the fuzzy sphere
converges to the sphere, but has the following additional features.
\begin{itemize}

\item $n^a$ and $N^a$ are bilinears in Grassmann variables.   This
should become irrelevant as $p \rightarrow\infty$ because these
combinations involve infinite sums of Grassmann bilinears and so all
of the vanishing relations in the Grassmann algebra only come in
very high order products.   The $L^2 $ sections involve sums of
finite monomials in these variables (generalized spherical
harmonics) with coefficients converging to zero rapidly with the
order of the monomial.

\item Conventionally one takes a limit of the fuzzy sphere which
produces a spherical geometry with finite radius.   Here we want to
make the radius infinite and obtain objects which depend only on the
conformal equivalence class of the geometry.   The conformal group
of the sphere is $SO(1,3)$, which is identified with the Lorentz
transformations.

\item As a consequence, wave functions will depend on a variable
$p \in R^+$, a continuous limit of the discrete $p$.   The
continuous $p$ will rescale under conformal transformations of the
sphere \cite{ashtekar}\cite{vn}.

\end{itemize}

If this conjecture is correct, then our limiting single particle
Hilbert space will be that of the massless chiral supermultiplet. We
will realize different particles in terms of disjoint $K \times K +
1$ blocks of the $N \times N + 1$ matrix $\psi_i^A$.  The ratios of
matrix sizes for different particles take all real positive values
in the limit, so we can parametrize the size of a given matrix by a
positive real number $p$\footnote{We are actually describing a limit
in which the matrices become elements of the hyperfinite
$II_{\infty}$ factor, as in \cite{vn}.} . The real linear
combinations of the operators $\psi_i^A$ and their adjoints converge
to linear functionals (operator valued measures) $S_a ({\bf \Omega},
p)$ on the space of measurable sections of the real spinor bundle
over the two sphere. The commutation relations are
$$ [S[f] , S[g] ]_+ = p f_a g_a ,$$ where $f_a ({\bf \Omega}) $ and
$g_a ({\bf \Omega})$ are any two sections.   If we choose
$q_a^{\alpha}$ to be the four real solutions of the CKS equation
(linear combinations of the solutions described above), then
$Q^{\alpha} = S[q^{\alpha}] $ satisfy the SUSY anti-commutation
relations for a single massless superparticle with momentum $p (1,
{\bf \Omega}) $ .

Our explicit construction leads directly to the chiral multiplet.
Bundles on the two sphere of charge $k$ ($2k$ is the power of the
positive chirality spinor bundle) are obtained from fuzzy modules of
$N \times N + 2k$ matrices.   We do not see how to obtain these
while simultaneously enforcing the requirement that the operators
$\psi$ converge to something that transforms in the spinor bundle.

When discussing representations of the massless SUSY algebra in four
dimensions, one is similarly led most naturally to the chiral
multiplet.   One simply appends a phase to the transformation law of
the states in order to describe higher spin massless multiplets.  It
is possible that one can construct a similar argument for the
present system, but we do not see how to obtain this from a natural
finite $N$ construction.

One way to do it is to insist that the pixel degrees of freedom
actually correspond to full $32$ component spinors, and satisfy a
version of the massless SUSY anticommutation relations of 11D SUGRA
in the presence of central charges.   Then one representation will
always contain the graviton.    Since we start from a holographic
description of the theory, it is reasonable to assume that it will
only be sensible as a dynamical theory, if it contains a graviton.
If there is no quick and dirty way, like that alluded to in the
preceding paragraph, to model higher helicities then we would find
that our formalism only makes sense if we model four dimensional
space-time as a compactification of String/M Theory.   Of course, a
lot more work needs to be done, before we could make such a
grandiose claim.

 Having dealt with single particle states, we turn to
multiparticle states. The basic idea is to consider block
decompositions of the full $N\times (N+1)$ matrix, where by the
previous argument, each individual block corresponds to a single
particle. We take the block sizes $1\ll p_i \ll N$ and take $N$ and
all $p_i$ to infinity, with $p_i / p_j$ fixed.

In particular, let's consider the following block decomposition
\eqn{hordecomp}{\psi = \pmatrix{1&2&\ldots &K \cr K&1&\ldots &K -1
\cr \ldots &\ldots &\ldots &\ldots \cr \ldots &\ldots &\ldots
&\ldots \cr 2&3&\ldots &1},} where $K \sim \sqrt{N}$.   We associate
the degrees of freedom labeled by a given integer $1 \leq p \leq K$
with a single independent horizon volume.   Note that, if we follow
the hint from our black hole discussion, and treat exchanges of
indices on $\psi$ as a gauge equivalence, then the different horizon
volumes are equivalent to each other. Furthermore, the different
blocks in a given horizon volume are indistinguishable, in the sense
that permutations of their order is just a relabeling.  As in Matrix
Theory\cite{bfss}, we will treat this as the gauge symmetry of
particle statistics.

We are free to vary the individual block sizes in a given horizon
volume, but if we want to maximize the entropy, with the proviso
that there be many individual particles, we should take the blocks
to be approximately square, with $K \sim \sqrt{N} $ rows.   If we do
this, all horizon volumes are indeed treated equally.   If we try to
increase an individual block size to be $\gg K$, then we
simultaneously squeeze out degrees of freedom in other horizons, and
constrain the allowed states of particles in our own horizon volume.
Thus, the idea that localized entropy in a given horizon volume is
\lq\lq borrowed" from the horizon, an idea which originates in the
black hole entropy formula, becomes quite explicit in this
construction. When we make such a large block, a description of the
system in terms of black holes becomes more appropriate.   We begin
to see the vague outlines of a unified description of black hole and
particle states, and their interactions\footnote{We have to admit
that it is rather too vague for our taste at the moment.}

The natural unit for this discrete momentum, is ${1\over R}$, the
minimal momentum that fits inside a horizon volume.  The maximal
momentum for particles in our maximal entropy configuration, with
block size $K$, is of order ${\sqrt{N}\over R} \sim \sqrt{{M_P}/R}$
in agreement with our field theory estimate.   However, unlike field
theory, our formalism allows us to take individual particle momenta
much larger than this, at the expense of making the momenta of other
particles smaller, or reducing the total number of particles.

Using only the degrees of freedom in a single horizon (corresponding
to a single integer label in our block decomposition), we can make a
maximal block size of order $N^{3/4}$, which would appear to give a
maximal momentum of order $ (RM_P)^{- 1/4} M_P $ (in the real world
this would be a few TeV).   However, this is not the only way to
make high momentum particles.   Our formalism describes particles by
the way in which they register on the holographic screen at
infinity.   As we will see in a moment, particles of higher
momentum, defined by full blocks of size $J \gg K$, have higher
angular resolution on the screen.   The full set of degrees of
freedom in a block of size $J$ describes superparticles whose
angular wave function can be roughly any one of the first $J$
spherical harmonics.   Thus, these operators can describe of order
$J$ particles, with the same absolute value of the momentum ${J\over
R} (1, {\bf\Omega})$.     If we only need to describe a few
particles with high momentum, we can use smaller blocks, but with
the wave functions \lq\lq locked together\rq\rq .   Thus, if we only
use $K$ blocks of size $K \sim \sqrt{N}$ but insist on states with
exactly the same angular wave function in each block, we are
describing the amplitude for a pixel in the detector on the
holoscreen to absorb momentum $M_P$.   From the point of view of
scattering theory this is interpreted as a single particle with
momentum $M_P$ .   Perhaps, in analogy with Matrix Theory, it should
be viewed as a bound state of the particles associated with
individual blocks. Note that, in contrast to Matrix theory, there is
no way to talk about particles at finite separation in the
kinematics we are discussing here. So the concept of a bound state
does not have an obvious meaning in this context.  In a similar
manner, we could try to describe particles with momentum up to $
(RM_P)^{1/2} M_P$ as bound states of $N^{3/2}$ blocks of size $\sim
1$.   These would be forced into a an almost unique angular wave
function, consisting of only the first few spherical harmonics.  The
description of particles in our formalism is thus flexible, and the
number of particles and their allowed four momentum wave functions
are constrained in a complicated and mutual way.

We end this section with a remark about the emergent $SU(2)$ group
of three dimensional rotations in the limiting asymptotically flat
theory, which we obtain as $(R M_P) \rightarrow \infty$.   It is not
the same as the $SU(2)$ group of the static observer, under which
the full fermionic matrix transforms as the $[N \otimes N + 1 ]$.
Our entire formalism was built on the hypothesis that the Poincar\'e
Hamiltonian $P_0$ was a very different operator from the static
Hamiltonian $H$.   Now we see that the same is true for the rotation
group.

\section{Finite N corrections}

If our mathematical conjecture about the limiting space of Grassman
wave functions is correct, then we have isolated the kinematic
degrees of freedom of particle physics from a formulation of the
quantum theory of dS space, which has a finite number of states. We
could then hope to get some insight about the finite c.c.
corrections to particle physics observables.  In particular, it is
plausible that the super-Poincar\'e commutation relation
\eqn{supoin}{[P_{\mu} , Q_{\alpha} ] = 0 , } (with $P_{\mu}$ {\it
defined}, as above, by the SUSY anticommutation relation) is not
valid at finite $N$.

In order to describe the spectrum of particles at a given mass scale
$m$, we should work with particles described as fermionic matrices
of size $(m R)$.   It is clear from the above discussion that unless
$ (m R) \gg 1$, we cannot hope to obtain a description which
approximates particles moving in flat space.   This remark shows
that {\it our formalism cannot describe gravitinos obeying the
classical SUGRA relation $m_{3/2} \sim {1\over R}$, which arises
from requiring that the c.c. not look fine tuned in the low energy
effective field theory sense.}   This argument alone cannot fix the
dependence of $m_{3/2}$ on $R$, but the most symmetric treatment of
particles in this system uses blocks of size $N^{1/2}$.   Thus one
might expect corrections to SUSY degeneracies of order $N^{- 1/2}$,
which suggests $m_{3/2} \sim (R/M_P)^{- 1/2} \sim \Lambda^{1/4}$, as
conjectured in \cite{tbfolly}.

\section{Conclusions}

We have presented a kinematic framework for the quantum theory of de
Sitter space, and identified configurations of the fundamental
variables which could represent both black hole and particle states.
Much remains to be done in order to develop this into a full blown
theory.   We list the most salient points:

\begin{itemize}

\item We must prove the conjecture that our finite dimensional particle Hilbert space
converges to the usual Fock space of superparticles.

\item We must understand how to describe compactified internal
dimensions and the spectrum of non-gravitational supermultiplets.
Indeed, in the limit $N \rightarrow \infty$ we expect our model to
have an S-matrix which is super-Poincar\'e invariant, and is likely to
be closely related to well understood string and M theory
constructions.   We might hope, {\it e.g.} that the limiting model
will have an approximate description as 11D SUGRA compactified on a
manifold of G2 holonomy with values of the moduli frozen at an R
symmetric point.   We need a kinematic description of such a
compactification in terms of fermionic matrices.

\item Most importantly, we need to formulate dynamical equations
which determine the scattering matrix and the object which
approximates it for finite $N$.

\end{itemize}

The first two of these desiderata seem within reach, while the third
remains somewhat mysterious.   One of the authors has suggested
possible avenues of attack on the dynamical problem in \cite{vn}.

\section{Appendix}

In this brief appendix, we indicate how to prove the conjectures we
made about random Hamiltonians.   These results were explained to us
by Mark Srednicki. We want to study random Hamiltonians whose
spectrum is bounded between $[0, E_b]$, where we will eventually
take $E_b$ to be of order the dS temperature.   Most studies refer
to Hamiltonians chosen from a Gaussian random ensemble.   We will
assume that similar results are valid for our case.

Given such a Hamiltonian, and in the limit of a large Hilbert space,
one can show that the time averaged expectation values of a class of
observables converge to the expectation values in a thermal
ensemble\cite{sred}. The necessary constraint on observables has to
do with their matrix elements in the basis where the Hamiltonian
matrix elements are Gaussian random variables.

The temperature of the thermal ensemble is related to the center of
the small energy band that is allowed in the eigenbasis expansion of
the initial state.   It is thus determined by the expectation value
of the energy in the initial state.

This expectation value is

$$\sum |a_i|^2 E_i .$$  We now want to average this over all
possible initial states.   The measure is the unitary invariant
measure on the unit sphere in our finite dimensional Hilbert space,
$$\delta (1 - \sum |a_i|^2 ) .$$   For large $N$ this can be
replaced by the Gaussian measure ${{e^{ - {1\over 2} \sum |a_i |^2
}}\over Z}.$   We can compute expectation values and fluctuations
using Wick's theorem. The expectation value is the average
eigenvalue of the Hamiltonian, and the fluctuations in this quantity
as we run over the ensemble of states is ${1\over N}$. Since the
energy scale is set by $E_b$ for any smooth measure with support on
the interval, this will also be the scale of the average
temperature. We choose $E_b$ so that the temperature is precisely
the dS temperature. Recall that in the dS case, $N$ is an enormously
large number.   For our own universe it would be $e^{10^{120}}$.

Thus, assuming that random Hamiltonians with a fixed upper and lower
bound thermalize generic states, we have proven the claims in the
text.
\section{Acknowledgments}

We would like to thank Willy Fischler and Lorenzo Mannelli for
contributing to the suite of ideas that formed the basis of this
paper.  Mark Srednicki helped us to understand the relation between
random Hamiltonians and thermalization. B.F. would like to thank the
organizers of the IV Simons Workshop and the Aspen Center for
Physics for hospitality during the course of this work. This research
was supported in part by DOE grant number DE-FG03-92ER40689.




  %




\end{document}